\documentclass[12pt]{article}
\usepackage{axodraw,epsfig}
\begin{document}

\newcommand{\be}[1]{\begin{equation}\label{#1}}
\newcommand{\beq}{\begin{equation}}
\newcommand{\ee}{\end{equation}}
\newcommand{\beqn}[1]{\begin{eqnarray}\label{#1}}
\newcommand{\eeqn}{\end{eqnarray}}
\newcommand{\bd}{\begin{displaymath}}
\newcommand{\ed}{\end{displaymath}}
\newcommand{\mat}[4]{\left(\begin{array}{cc}{#1}&{#2}\\{#3}&{#4}
\end{array}\right)}
\newcommand{\matr}[9]{\left(\begin{array}{ccc}{#1}&{#2}&{#3}\\
{#4}&{#5}&{#6}\\{#7}&{#8}&{#9}\end{array}\right)}
\newcommand{\matrr}[6]{\left(\begin{array}{cc}{#1}&{#2}\\
{#3}&{#4}\\{#5}&{#6}\end{array}\right)}
\def\lsim{\raise0.3ex\hbox{$\;<$\kern-0.75em\raise-1.1ex
\hbox{$\sim\;$}}}
\def\gsim{\raise0.3ex\hbox{$\;>$\kern-0.75em\raise-1.1ex
\hbox{$\sim\;$}}}
\def\abs#1{\left| #1\right|}
\def\simlt{\mathrel{\lower2.5pt\vbox{\lineskip=0pt\baselineskip=0pt
           \hbox{$<$}\hbox{$\sim$}}}}
\def\simgt{\mathrel{\lower2.5pt\vbox{\lineskip=0pt\baselineskip=0pt
           \hbox{$>$}\hbox{$\sim$}}}}
\def\unity{{\hbox{1\kern-.8mm l}}}
\def\epr{E^\prime}
\newcommand{\al}{\alpha}
\def\16p{16\pi^2}
\def\ga{\gamma}
\def\Ga{\Gamma}
\def\la{\lambda}
\def\La{\Lambda}
\def\al{\alpha}
\newcommand{\ov}{\overline}
\renewcommand{\to}{\rightarrow}
\renewcommand{\vec}[1]{\mbox{\boldmath$#1$}}
\def\mcirc{{\stackrel{o}{m}}}
\newcommand{\tanb}{\tan\beta}
\def\dfrac#1#2{{\displaystyle\frac{#1}{#2}}}

\begin{titlepage}

\begin{flushright}
{DFPD-03/TH/11} \\

\end{flushright}

\vspace{2.0cm}

\begin{center}

{\Large  \bf 
  Lepton Flavour Violating Decays  \\
\vspace{0.4cm}
of Supersymmetric  Higgs Bosons }

\vspace{0.7cm}

{\large Andrea Brignole\footnote{ E-mail address: andrea.brignole@pd.infn.it} 
and Anna Rossi\footnote{ E-mail address: anna.rossi@pd.infn.it}}

\vspace{10mm}

{\it  Dipartimento di Fisica `G.~Galilei',  
Universit\`a di Padova and \\
\vspace{0.1cm}
INFN, Sezione di Padova,
Via Marzolo 8, I-35131 Padua, Italy}

\end{center}

\vspace{8mm}

\begin{abstract}
We compute the lepton flavour violating couplings of 
Higgs bosons in the Minimal Supersymmetric Standard Model, 
and show that they can induce the decays $(h^0,H^0,A^0) \to \mu \tau$ 
at non-negligible rates, for large $\tan\beta$ and sizeable   
smuon-stau mixing.
We also discuss the prospects for detecting such decays 
at LHC and other colliders, as well as the correlation 
with other flavour violating processes, such as    
$\tau \to \mu \gamma$ and $\tau \to 3 \mu$.
\end{abstract}

\end{titlepage}

\setcounter{footnote}{0}

\section{Introduction}

The recent important indications of neutrino oscillations  \cite{nus}
reveal  that flavour violation  also occurs in the lepton sector and 
further motivate the search for alternative signals of 
lepton flavour violation (LFV).
The Minimal Supersymmetric  extension of the Standard Model (MSSM) is a  
natural framework where several such signals  could be significant, 
provided the mass matrices of the leptons and of the sleptons
are not aligned.
Well known examples are  the LFV radiative decays of charged leptons, 
$\mu\to e \ga, \tau \to  \mu \ga, \tau \to  e \ga $. 
In this Letter we would like to explore another class of 
such processes, namely  the LFV decays 
of the neutral Higgs bosons ($h^0, H^0, A^0$). 
An important feature of these decays is that the corresponding 
amplitudes do not vanish in the limit of very heavy superpartners,
since the leading contributions  
are induced by  dimension-four effective operators,  
at variance with the case of radiative decays.

Related investigations on flavour violating Higgs
couplings in the MSSM framework have mainly focused
on processes with {\it virtual} Higgs exchange 
(see e.g. \cite{QFV,BK,sher1,DER}) and regard either quark or lepton 
flavour violation.
The decays of {\it physical} Higgs bosons
into fermion pairs have been explored in the case of 
quark flavour violation in the MSSM \cite{CHT,demir}, whereas in the case
of lepton flavour violation existing studies \cite{sher,gen} have mainly
used phenomenological parametrizations of the LFV couplings\footnote{
An attempt to study LFV Higgs decays in the MSSM can be found 
in \cite{DC}.  However, we believe that in this work the 
Higgs couplings to the sleptons have not been properly identified.}.

This Letter is organized as follows. 
In Section 2 we present the effective LFV 
Higgs couplings  in the 
MSSM framework, focusing on the second and third lepton  generations.
We explicitly compute  the one-loop contributions 
to those couplings and   the branching ratios 
of the decays $(h^0, H^0, A^0) \to \mu \tau$. New results 
on flavour conserving Higgs couplings are also presented. 
In Section 3 we give a numerical discussion on the LFV Higgs 
couplings and branching ratios, and also discuss the 
prospects at future colliders. Finally in Section 4 
we  comment on the correlation of the LFV Higgs decays 
with other LFV processes, such as $\tau\to \mu\ga$ and 
$\tau \to 3 \mu$, and  summarize our results.
 
\section{Higgs-muon-tau effective interactions}

The MSSM contains two Higgs doublets $H_1$ and $H_2$, with 
opposite hypercharges. Down-type fermions, which only
couple to $H_1$ at the tree level, also couple to $H_2$ 
after the inclusion of radiative corrections \cite{bottom}.
In particular, for the charged leptons of second and third 
generations the tree-level couplings read as  
\be{Ltree}
{\cal L} = -  Y_\mu H^0_1 \mu^c \mu  -Y_\tau H^0_1 \tau^c \tau + {\rm h.c.} , 
\ee
where $H^0_1$  is the neutral component of  $H_1$  and 
$Y_\mu, Y_\tau$ are the Yukawa coupling constants\footnote{
We adopt  two-component spinor notation, so 
$\mu$ and $\tau$ ($\bar{\mu}^c$ and $\bar{\tau}^c$) are the left-handed 
(right-handed) components of the muon and tau fields, respectively.
Throughout our discussion we assume CP conservation and therefore 
all the dimensionless as well as dimensionful parameters are  
taken to be real.}. 
Also the leading effective interactions with $H_2$, which arise
once superpartners are integrated out, are described by
dimension-four operators. These can be either
flavour conserving (FC):
\be{lefffc}
\Delta{\cal L}_{FC} = 
- (Y_\mu \Delta_{\mu} + Y_\tau \Delta'_\mu ) {H}^{0*}_2 
\mu^c \mu
-Y_\tau \Delta_{\tau} {H}^{0*}_2 \tau^c \tau 
 +{\rm h.c.}  , 
\ee
or flavour violating (FV):
\be{leff}
\Delta{\cal L}_{FV} = 
-Y_\tau \Delta_{L} {H}^{0*}_2 \tau^c \mu
-Y_\tau \Delta_{R} {H}^{0*}_2 \mu^c \tau 
 +{\rm h.c.}  , 
\ee
where   $\Delta_\mu, \Delta'_\mu, \Delta_\tau$ and $\Delta_L, \Delta_R$  
are  dimensionless functions of 
the MSSM mass parameters, to be described below. 
In eqs.~(\ref{lefffc}) and (\ref{leff})  we have only retained 
the dominant terms, proportional to $Y_\tau$, besides the first 
term in $\Delta {\cal L}_{FC}$ proportional to $Y_\mu$. 
In the following we are mostly interested in the effects induced 
by the terms in (\ref{leff}).
In the mass-eigenstate basis for both leptons and Higgs bosons, 
the FV couplings read as:
\be{LFV}
{\cal L}_{FV} = - \frac{Y_\tau}{\sqrt2 \cos\beta} 
( \Delta_L ~\tau^c \mu +\Delta_R~ \mu^c \tau )~ 
[ h^0 \cos(\beta -\alpha)
-H^0 \sin(\beta -\alpha) - {\rm i} A^0] 
+ {\rm h.c.}  , 
\ee
where $\tan\beta\! = \!\langle H^0_2\rangle/\langle H^0_1\rangle$, 
$\alpha$ is the  mixing angle in the CP-even Higgs sector 
[$\sqrt2 {\rm Re}(H^0_1 -  \langle H^0_1\rangle) = 
 H^0 \cos\al -  h^0 \sin\al$, 
$ ~ \sqrt2 {\rm Re}(H^0_2 -  \langle H^0_2\rangle) 
=  H^0\sin\al +  h^0 \cos\al$] and $A^0$ is the physical CP-odd 
Higgs field. The expression in (\ref{LFV}) holds up to 
${\cal O}(\Delta_\tau \tan\beta)$ corrections, which arise from 
eq.~(\ref{lefffc}) and can be
${\cal O}(10\%)$  for large $\tan\beta$. For our purposes it is not 
compelling to include and  resum such higher-order 
($\tan\beta$-enhanced) terms.

The effective couplings (\ref{LFV}) contribute to LFV low-energy processes, 
such as the decay $\tau\to 3 \mu$ and other ones,  through  
Higgs boson  exchange \cite{BK,sher1,DER}. 
We will comment later on $\tau\to 3 \mu$.
Here we are interested in a more direct implication of those LFV 
couplings, i.e.  the decays $\Phi^0\to \mu^{\pm}\tau^{\mp}$ 
where $\Phi^0= h^0, H^0, A^0$.
It is straightforward to compute the branching ratios 
$BR(\Phi^0\to \mu^+\tau^-)= BR(\Phi^0\to \mu^-\tau^+)$, 
 and  it is convenient 
to relate them to those of the 
flavour conserving decays  $\Phi^0\to \tau^+ \tau^-$:
\be{rphi}
{ BR(\Phi^0\to \mu^+\tau^-)}
=  \tan^2\beta~ (|\Delta_L|^2 + |\Delta_R|^2) 
 ~ C_\Phi~ { BR(\Phi^0\to \tau^+\tau^-)}  \, ,
\ee
where the $C_\Phi$ coefficients are:
\be{cphi}
C_h = \left[\frac{\cos(\beta - \al)}{\sin\al}\right]^2, ~~~~
C_H = \left[\frac{\sin(\beta - \al)}{\cos\al}\right]^2, ~~~~
C_A = 1 .
\ee
Since non-negligible effects can  only arise in the 
large $\tan\beta$ limit, in eq.~(\ref{rphi})    
we have approximated $1/\cos^2\beta \simeq \tan^2\beta$.

We now present explicit expressions for the quantities $\Delta_{L}$ and 
$\Delta_R$, i.e. the coefficients of the dimension-four operators
in (\ref{leff}). The relevant one-loop diagrams, which 
involve the exchange of sleptons, gauginos and 
Higgsinos, are shown in Fig.~1. The diagrammatic computation
is consistently performed in the gauge symmetry limit,
at zero external momentum\footnote{In particular,  
the only Higgs insertion we consider is that explicitly depicted in 
the diagrams. Further Higgs insertions or momentum dependent effects
correspond to higher dimension operators and give subleading
corrections to $\Phi^0\to \mu \tau$, in the limit of heavy 
superpartners and large $\tan\beta$.}.
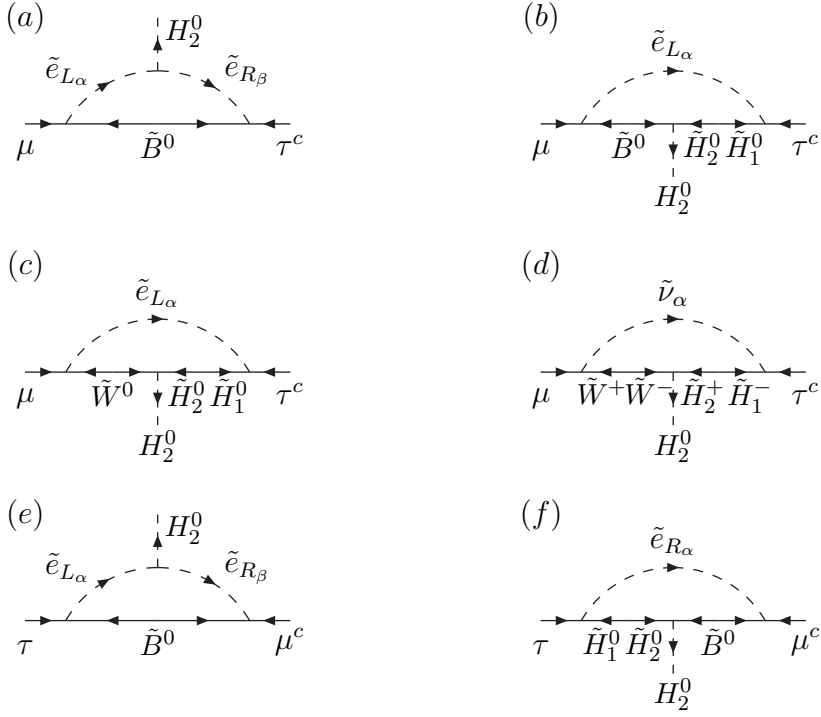
\begin{figure}[t]
\vspace{0.4cm}
\begin{center}
\begin{picture}(120,80)(-60,-40)
\ArrowLine(-50,0)(-34,0)
\ArrowLine(0,0)(-34,0)
\ArrowLine(0,0)(34,0)
\ArrowLine(50,0)(34,0)
\DashArrowArcn(0,-20)(40,150,90){4}
\DashArrowArcn(0,-20)(40,90,30){4}
\DashArrowLine(0,20)(0,40){3}
\Text(-50,-10)[]{$\mu$}
\Text(50,-8)[]{$\tau^c$}
\Text(0,-8)[]{\small $\tilde{B}^0$}
\Text(10,35)[]{\small $H_2^0$}
\Text(-34,20)[]{$\tilde{e}_{L_\alpha}$}
\Text(34,20)[]{$\tilde{e}_{R_\beta}$}
\Text(-50,40)[]{$(a)$}
\end{picture}
\hglue 2.5cm
\begin{picture}(120,80)(-60,-40)
\ArrowLine(-50,0)(-34,0)
\ArrowLine(-17,0)(-34,0)
\ArrowLine(-17,0)(0,0)
\ArrowLine(17,0)(0,0)
\ArrowLine(17,0)(34,0)
\ArrowLine(50,0)(34,0)
\DashArrowArcn(0,-20)(40,150,30){4}
\DashArrowLine(0,0)(0,-20){3}
\Text(-50,-10)[]{$\mu$}
\Text(50,-8)[]{$\tau^c$}
\Text(27,-9)[]{\small $\tilde{H}_1^0$}
\Text(11,-9)[]{\small $\tilde{H}_2^0$}
\Text(-17,-8)[]{\small $\tilde{B}^0$}
\Text(0,-28)[]{\small $H_2^0$}
\Text(0,30)[]{$\tilde{e}_{L_\alpha}$}
\Text(-50,40)[]{$(b)$}
\end{picture}
\end{center}
\begin{center}
\begin{picture}(120,80)(-60,-40)
\ArrowLine(-50,0)(-34,0)
\ArrowLine(-17,0)(-34,0)
\ArrowLine(-17,0)(0,0)
\ArrowLine(17,0)(0,0)
\ArrowLine(17,0)(34,0)
\ArrowLine(50,0)(34,0)
\DashArrowArcn(0,-20)(40,150,30){4}
\DashArrowLine(0,0)(0,-20){3}
\Text(-50,-10)[]{$\mu$}
\Text(50,-8)[]{$\tau^c$}
\Text(27,-9)[]{\small $\tilde{H}_1^0$}
\Text(11,-9)[]{\small $\tilde{H}_2^0$}
\Text(-17,-8)[]{\small $\tilde{W}^0$}
\Text(0,-28)[]{\small $H_2^0$}
\Text(0,30)[]{$\tilde{e}_{L_\alpha}$}
\Text(-50,40)[]{$(c)$}
\end{picture}
\hglue 2.5cm
\begin{picture}(120,80)(-60,-40)
\ArrowLine(-50,0)(-34,0)
\ArrowLine(-17,0)(-34,0)
\ArrowLine(-17,0)(0,0)
\ArrowLine(17,0)(0,0)
\ArrowLine(17,0)(34,0)
\ArrowLine(50,0)(34,0)
\DashArrowArcn(0,-20)(40,150,30){4}
\DashArrowLine(0,0)(0,-20){3}
\Text(-50,-10)[]{$\mu$}
\Text(50,-8)[]{$\tau^c$}
\Text(29,-9)[]{\small $\tilde{H}_1^-$}
\Text(11,-9)[]{\small $\tilde{H}_2^+$}
\Text(-9,-7)[]{\small $\tilde{W}^-$}
\Text(-27,-7)[]{\small $\tilde{W}^+$}
\Text(0,-28)[]{\small $H_2^0$}
\Text(0,30)[]{$\tilde{\nu}_\alpha$}
\Text(-50,40)[]{$(d)$}
\end{picture}
\end{center}
\begin{center}
\begin{picture}(120,80)(-60,-40)
\ArrowLine(-50,0)(-34,0)
\ArrowLine(0,0)(-34,0)
\ArrowLine(0,0)(34,0)
\ArrowLine(50,0)(34,0)
\DashArrowArcn(0,-20)(40,150,90){4}
\DashArrowArcn(0,-20)(40,90,30){4}
\DashArrowLine(0,20)(0,40){3}
\Text(-50,-10)[]{$\tau$}
\Text(50,-8)[]{$\mu^c$}
\Text(0,-8)[]{\small $\tilde{B}^0$}
\Text(10,35)[]{\small $H_2^0$}
\Text(-34,20)[]{$\tilde{e}_{L_\alpha}$}
\Text(34,20)[]{$\tilde{e}_{R_\beta}$}
\Text(-50,40)[]{$(e)$}
\end{picture}
\hglue 2.5cm
\begin{picture}(120,80)(-60,-40)
\ArrowLine(-50,0)(-34,0)
\ArrowLine(-17,0)(-34,0)
\ArrowLine(-17,0)(0,0)
\ArrowLine(17,0)(0,0)
\ArrowLine(17,0)(34,0)
\ArrowLine(50,0)(34,0)
\DashArrowArcn(0,-20)(40,150,30){4}
\DashArrowLine(0,0)(0,-20){3}
\Text(-50,-10)[]{$\tau$}
\Text(50,-8)[]{$\mu^c$}
\Text(-27,-9)[]{\small $\tilde{H}_1^0$}
\Text(-11,-9)[]{\small $\tilde{H}_2^0$}
\Text(17,-8)[]{\small $\tilde{B}^0$}
\Text(0,-28)[]{\small $H_2^0$}
\Text(0,30)[]{$\tilde{e}_{R_\alpha}$}
\Text(-50,40)[]{$(f)$}
\end{picture}
\end{center}
\vspace{-0.3cm}
\caption{\small Diagrams that contribute to $\Delta_L$
[(a),(b), (c),(d)] and to $\Delta_R$ [(e),(f)].} 
\label{f1}
\end{figure}
In the superfield basis in which the charged lepton mass matrix 
is diagonal,       
the mass matrices of the left-handed and right-handed sleptons read:
\be{matri}
\tilde{{\cal M}}^2_L =
\pmatrix{
\tilde{m}^2_{L \mu \mu} & \tilde{m}^2_{L \mu \tau} \cr
\tilde{m}^2_{L \mu \tau} & \tilde{m}^2_{L \tau \tau} }, \, \,~~~~
\tilde{{\cal M}}^2_R = \pmatrix{
\tilde{m}^2_{R \mu \mu} & \tilde{m}^2_{R \mu \tau} \cr
\tilde{m}^2_{R \mu \tau} & \tilde{m}^2_{R \tau \tau} } .
\end{equation}
We are interested in scenarios with 
large LFV, either in $\tilde{{\cal M}}^2_L$ 
[$({\rm LFV})_L$] or in $\tilde{{\cal M}}^2_R$ [$({\rm LFV})_R$].  
Large $({\rm LFV})_L$ means that $\tilde{m}^2_{L \mu \tau}$ 
is comparable to $\tilde{m}^2_{L \mu \mu}$ and  $\tilde{m}^2_{L \tau\tau}$. 
Similarly, large  $({\rm LFV})_R$ means that $\tilde{m}^2_{R \mu \tau}$ 
is comparable to $\tilde{m}^2_{R \mu \mu}$ and  $\tilde{m}^2_{R \tau\tau}$. 
The flavour states $\tilde{L}_\mu = 
(\tilde{\nu}_\mu , \tilde{\mu}_L)^T, 
\tilde{L}_\tau = 
(\tilde{\nu}_\tau , \tilde{\tau}_L)^T$ are related to the 
$\tilde{{\cal M}}^2_L$ eigenstates 
$\tilde{L}_2 = 
(\tilde{\nu}_2 , \tilde{e}_{L_2})^T, 
\tilde{L}_3 = 
(\tilde{\nu}_3 , \tilde{e}_{L_3})^T$ 
by the relations $\tilde{L}_\mu = c_L \tilde{L}_2 - s_L \tilde{L}_3, ~
\tilde{L}_\tau = s_L \tilde{L}_2 + c_L \tilde{L}_3$. 
Analogous relations hold for the right-handed 
sleptons:   $\tilde{\mu}_{R} = c_R \tilde{e}_{R_2} - s_R \tilde{e}_{R_3}, ~
\tilde{\tau}_{R} = s_R \tilde{e}_{R_2} + c_R \tilde{e}_{R_3}$, 
where $ \tilde{e}_{R_2}$ and $ \tilde{e}_{R_3}$ are the eigenstates of 
$\tilde{{\cal M}}^2_R$.
The mixing parameters satisfy the following relations:
\be{mix}
s_L c_L = 
\frac{\tilde{m}^2_{L \mu \tau}}{\tilde{m}^2_{L_2}-\tilde{m}^2_{L_3} }~ , 
~~~~~~~s_R c_R = 
\frac{\tilde{m}^2_{R \mu \tau}}{\tilde{m}^2_{R_2}-\tilde{m}^2_{R_3} } ~, 
\ee
where $\tilde{m}^2_{L_\al}$ and $\tilde{m}^2_{R_\al}$ ($\al =2,3$) are 
the eigenvalues of $\tilde{{\cal M}}^2_L$ and $\tilde{{\cal M}}^2_R$, 
respectively.
The other relevant parameters are  the Bino ($\tilde{B}$) mass $M_1$, 
the Wino ($\tilde{W}^0, \tilde{W}^{\pm}$) mass $M_2$ and the 
$\mu$ parameter. The latter appears in the Higgsino mass terms 
$-\mu (\tilde{H}^0_1 
\tilde{H}^0_2 - \tilde{H}^-_1 
\tilde{H}^+_2) +{\rm h.c.}$ and in the cubic interaction $-Y_\tau \mu 
H^{0*}_2 \tilde{\tau}^*_R \tilde{\tau}_L   +{\rm h.c.}$. 
The explicit evaluation of the diagrams gives for $\Delta_L$:
\be{deltaL}
\Delta_L = \Delta^{(a)}_L + \Delta^{(b)}_L+\Delta^{(c)}_L 
+ \Delta^{(d)}_L ,
\ee
\beqn{deltaL2}
\Delta^{(a)}_L &=& - \frac{g'^2}{16 \pi^2} \mu M_1 s_L c_L\left[
s^2_R \left(
{\rm I}(M^2_1, \tilde{m}^2_{R_2}, \tilde{m}^2_{L_2})  -
{\rm I}(M^2_1, \tilde{m}^2_{R_2}, \tilde{m}^2_{L_3})\right)
\right. \nonumber \\ 
&&  \left. +c^2_R\left( {\rm I}(M^2_1, \tilde{m}^2_{R_3}, 
\tilde{m}^2_{L_2}) - 
{\rm I}(M^2_1, \tilde{m}^2_{R_3}, \tilde{m}^2_{L_3}) \right)\right]
, \nonumber\\
 \Delta^{(b)}_L &=& - \frac{g'^2}{32\pi^2} \mu M_1 s_L c_L\left[
{\rm I}(M^2_1,\mu^2, \tilde{m}^2_{L_2}) - 
{\rm I}(M^2_1,\mu^2, \tilde{m}^2_{L_3}) 
\right]  , \nonumber\\
 \Delta^{(c)}_L &=& \frac{g^2}{32 \pi^2} \mu M_2 s_L c_L\left[
{\rm I}(M^2_2,\mu^2, \tilde{m}^2_{L_2}) - 
{\rm I}(M^2_2,\mu^2, \tilde{m}^2_{L_3}) \right]  , \nonumber\\
\Delta^{(d)}_L &=& \frac{g^2}{16 \pi^2} \mu M_2 s_L c_L\left[
{\rm I}(M^2_2,\mu^2, \tilde{m}^2_{L_2}) - 
{\rm I}(M^2_2,\mu^2, \tilde{m}^2_{L_3}) \right]  , 
\eeqn
and for $\Delta_R$:
\be{deltaR}
\Delta_R = \Delta^{(e)}_R + \Delta^{(f)}_R , 
\ee
\beqn{deltaR2}
\Delta^{(e)}_R &=& - \frac{g'^2}{16 \pi^2} \mu M_1 s_R c_R\left[
s^2_L \left({\rm I}(M^2_1, \tilde{m}^2_{L_2}, \tilde{m}^2_{R_2}) 
-{\rm I}(M^2_1, \tilde{m}^2_{L_2}, \tilde{m}^2_{R_3})\right) 
 \right.\nonumber \\ 
&& \left.
+ c^2_L \left({\rm I}(M^2_1, \tilde{m}^2_{L_3},\tilde{m}^2_{R_2}) - 
{\rm I}(M^2_1, \tilde{m}^2_{L_3},\tilde{m}^2_{R_3})\right)
\right] , \nonumber\\
 \Delta^{(f)}_R &=&  \frac{g'^2}{16 \pi^2} \mu M_1 s_R c_R\left[
{\rm I}(M^2_1,\mu^2, \tilde{m}^2_{R_2}) - 
{\rm I}(M^2_1,\mu^2, \tilde{m}^2_{R_3})\right]  , 
\eeqn
The function ${\rm I}$, which has mass dimension $-2$,
is the standard three-point one-loop integral:
\be{In}
{\rm I}(x,y,z)= \frac{ xy \log\frac{x}{y}  +yz \log\frac{y}{z} 
+ zx \log\frac{z}{x}}{ (x-y) (z-y)(z-x)} .
\ee 
Our  results for the LFV diagrams  in Fig.~\ref{f1}  
can be compared with similar ones presented  in \cite{BK,DER}.
However, one notices some 
differences in those works: {\it i}) there LFV effects were
treated at linear order, through the mass insertion approximation; 
{\it ii}) only LFV in the left-handed sleptons was considered, since  
LFV was related to the seesaw generation of neutrino masses; 
{\it iii}) the relative signs between the $\tilde{B}$ diagram and 
gaugino-Higgsino diagrams differ from ours. 
This sign is crucial to correctly determine the interference effects,
as we will see below. Notice that such a sign discrepancy  
does not depend on the fact that we use a different 
sign convention  for the $\mu$ parameter. 

For the sake of completeness we also present the expressions 
of the FC parameters $\Delta_\mu, \Delta'_\mu, \Delta_\tau$, which are 
relevant for establishing the relations between the lepton masses 
($m_\mu$, $m_\tau$)  and the 
 corresponding Yukawa couplings.
Such quantities are induced by diagrams analogous 
to those in Fig.~1 but with the same 
flavour in the external fermion lines (either muon or tau flavour):
\beqn{deltamu}
\Delta_\mu & =&
- \frac{g'^2}{16 \pi^2} \mu M_1\left[
c^2_L c^2_R  
{\rm I}(M^2_1, \tilde{m}^2_{L_2}, \tilde{m}^2_{R_2}) 
+c^2_L s^2_R  
{\rm I}(M^2_1, \tilde{m}^2_{L_2}, \tilde{m}^2_{R_3}) \right. \nonumber \\
&& \left. 
 +s^2_L c^2_R  
{\rm I}(M^2_1, \tilde{m}^2_{L_3}, \tilde{m}^2_{R_2}) 
+s^2_L s^2_R  
{\rm I}(M^2_1, \tilde{m}^2_{L_3}, \tilde{m}^2_{R_3})
+\frac12 
c^2_L {\rm I}(M^2_1,\mu^2, \tilde{m}^2_{L_2}) \right.  \nonumber\\
&& \left. + \frac12 
s^2_L {\rm I}(M^2_1,\mu^2, \tilde{m}^2_{L_3})  
- c^2_R {\rm I}(M^2_1,\mu^2, \tilde{m}^2_{R_2})
- s^2_R {\rm I}(M^2_1,\mu^2, \tilde{m}^2_{R_3}) \right] \nonumber \\
&&
+\frac{3g^2}{32 \pi^2}\mu M_2 \left[c^2_L 
{\rm I}(M^2_2,\mu^2, \tilde{m}^2_{L_2})+ 
s^2_L 
{\rm I}(M^2_2,\mu^2, \tilde{m}^2_{L_3})\right]  , \\
\Delta'_\mu &= &
- \frac{g'^2}{16 \pi^2} \mu M_1 s_L c_L s_R c_R\left[
{\rm I}(M^2_1, \tilde{m}^2_{L_2}, \tilde{m}^2_{R_2}) 
-{\rm I}(M^2_1, \tilde{m}^2_{L_2}, \tilde{m}^2_{R_3}) \right. \nonumber \\
&& \left. -
{\rm I}(M^2_1, \tilde{m}^2_{L_3}, \tilde{m}^2_{R_2})+
{\rm I}(M^2_1, \tilde{m}^2_{L_3}, \tilde{m}^2_{R_3}) \right] , \\
\Delta_\tau & =&
- \frac{g'^2}{16 \pi^2} \mu M_1\left[
s^2_L s^2_R  
{\rm I}(M^2_1, \tilde{m}^2_{L_2}, \tilde{m}^2_{R_2}) 
+s^2_L c^2_R  
{\rm I}(M^2_1, \tilde{m}^2_{L_2}, \tilde{m}^2_{R_3}) \right. \nonumber \\
&& \left. 
 +c^2_L s^2_R  
{\rm I}(M^2_1, \tilde{m}^2_{L_3}, \tilde{m}^2_{R_2}) 
+c^2_L c^2_R  
{\rm I}(M^2_1, \tilde{m}^2_{L_3}, \tilde{m}^2_{R_3})
+\frac12 
s^2_L {\rm I}(M^2_1,\mu^2, \tilde{m}^2_{L_2}) \right.  \nonumber\\
&& \left. + \frac12 
c^2_L {\rm I}(M^2_1,\mu^2, \tilde{m}^2_{L_3})  
- s^2_R {\rm I}(M^2_1,\mu^2, \tilde{m}^2_{R_2}) 
- c^2_R {\rm I}(M^2_1,\mu^2, \tilde{m}^2_{R_3}) \right] \nonumber \\
&&
+\frac{3g^2}{32 \pi^2}\mu M_2 \left[s^2_L 
{\rm I}(M^2_2,\mu^2, \tilde{m}^2_{L_2})+ 
c^2_L 
{\rm I}(M^2_2,\mu^2, \tilde{m}^2_{L_3})\right]  .
\eeqn
These formulas are quite general as they include possible 
LFV in the slepton mass matrices. 
By setting $c_L =c_R=1$, $s_L=s_R=0$  
one can easily 
recover for $\Delta_\mu$ and $\Delta_\tau$ the corresponding cases\footnote{
In this limit of vanishing LFV, different expressions  for $\Delta_\tau$ 
can be found  
in the literature \cite{all,GHP,BK}, and some discrepancies 
exist among them. 
Our result for $\Delta_\tau$ is consistent with that in 
\cite{GHP},  taking into account that we use 
an opposite sign convention for the $\mu$ parameter and 
include left-right slepton mixing at linear order. 
To our knowledge no explicit expression for $\Delta_\mu$ 
or $\Delta'_\mu$ appears in the literature. In principle $\Delta_\mu$ 
can be distinct from $\Delta_\tau$.}   
without LFV, whereas $\Delta'_\mu$ vanishes as this term requires 
both   $({\rm LFV})_L$ and  $({\rm LFV})_R$.
Incidentally, notice 
that  $\Delta'_\mu$ in  eq.~(\ref{lefffc}) is multiplied by 
$Y_\tau$. 
Thereby, if $({\rm LFV})_L$ and  $({\rm LFV})_R$ are both large, 
the relation between the muon mass and Yukawa coupling  
could receive large  
($\tanb$ and  $Y_\tau/Y_\mu$ enhanced) 
corrections\footnote{
In this limit of large $({\rm LFV})_L$ {\it and}  $({\rm LFV})_R$, 
 analogous enhancement effects  also 
appear in the muon magnetic and electric dipole operators, 
see e.g. \cite{g2}.
For similar enhancement effects in the relation between
quark masses and Yukawa couplings, see e.g. \cite{demir}.}, 
and the ratios  
$BR(\Phi^0\to \mu^+ \mu^-)/BR(\Phi^0\to \tau^+ \tau^-)$ 
could differ significantly from the tree 
level expectation $(m_\mu/m_\tau)^2$.
However, having simultaneously large $({\rm LFV})_L$ 
and  $({\rm LFV})_R$ does not seem very natural
if the smallness of $m_\mu/m_\tau$ is related to 
an underlying supersymmetric flavour symmetry.

\section{Numerical results and implications at colliders}

Now we give some numerical examples to appreciate 
the size of the effects we are discussing. 
For definiteness, we discuss separately the case of large $({\rm LFV})_L$,  
with negligible $({\rm LFV})_R$,  and the complementary case of large 
$({\rm LFV})_R$,  with negligible $({\rm LFV})_L$. 
Let us redefine  in (\ref{matri}) 
$\tilde{m}^2_{L \tau \tau} \equiv \tilde{m}^2_L$ and 
$\tilde{m}^2_{R \tau \tau} \equiv \tilde{m}^2_R$.  
As a representative case of 
large $({\rm LFV})_L$, we choose
 $ \tilde{m}^2_{L \mu \mu}= \tilde{m}^2_L$ and 
$\tilde{m}^2_{L \mu \tau} = 0.8 \cdot \tilde{m}^2_L$, while 
$\tilde{m}^2_{R \mu \tau} \sim 0$. 
Analogously, for the case of large $({\rm LFV})_R$ 
we  choose $ \tilde{m}^2_{R \mu \mu}=
\tilde{m}^2_R$ and 
$\tilde{m}^2_{R \mu \tau} = 0.8 \cdot \tilde{m}^2_R$, while 
$\tilde{m}^2_{L \mu \tau} \sim 0$.
We show the quantity $|50 \Delta_L|^2$  
as a function of $|\mu|/\tilde{m}_L$ in Fig.~\ref{f2} and   
 $|50 \Delta_R|^2$ as a function of $|\mu|/\tilde{m}_R$ 
in Fig.~\ref{f3}, for fixed values of other mass ratios.  
We have inserted a factor 50 to make it easier 
the numerical estimate  of eq.~(\ref{rphi}) for the reference case of 
$\tan\beta=50$.
The curves depicted 
\begin{figure}[p]
\vskip -1.cm
\hglue 2.5cm
\epsfig{file=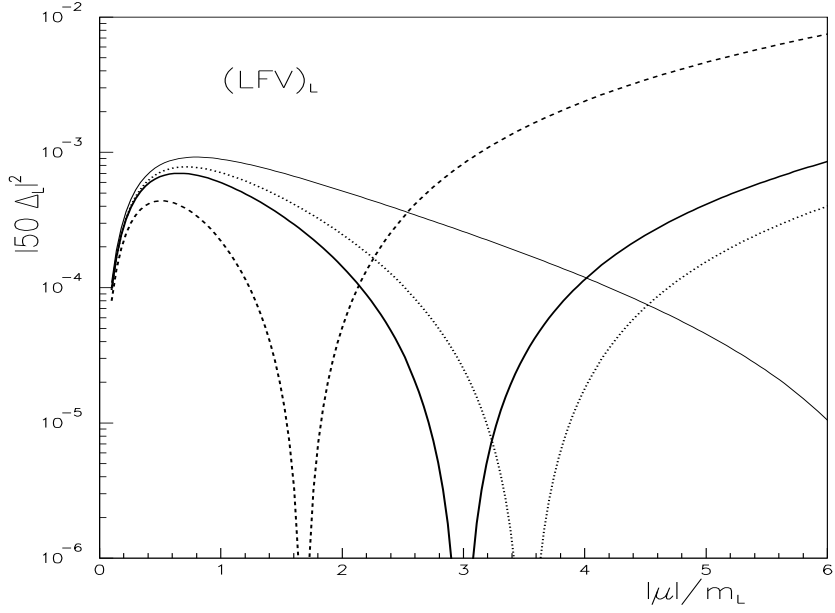,height=9.cm,width= 12.1cm}
\vskip -0.2cm
\caption{\small 
The quantity   $|50 \Delta_L|^2$ as a function  of $|\mu|/\tilde{m}_L$, 
for $\tilde{m}^2_{L \mu \tau} = 0.8 \cdot \tilde{m}^2_L$ 
 and four choices of 
the other relevant mass ratios:
1) $M_1= M_2 =\tilde{m}_R= \tilde{m}_L$ (solid line); 
2) $M_1= \tilde{m}_L/3$, $M_2 =\tilde{m}_R= \tilde{m}_L$ (dotted line); 
3) $M_1=M_2 = \tilde{m}_L$, $ \tilde{m}_R= \tilde{m}_L/3$ (dashed line); 
4) $M_1=M_2 = \tilde{m}_L$, $ \tilde{m}_R= 3\tilde{m}_L$ (thin solid line).
}
\label{f2}
\vskip -0.1cm
\end{figure}
\begin{figure}[p]
\vskip -1.cm
\hglue 2.5cm
\epsfig{file=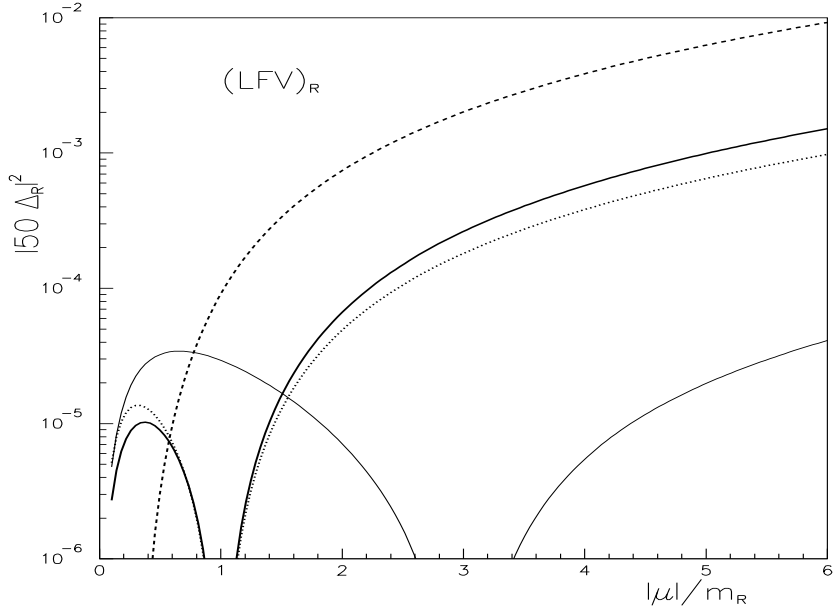,height=9.cm,width= 12.1cm}
\vskip -0.2cm
\caption{\small 
The quantity   $|50 \Delta_R|^2$  as a function 
of $|\mu|/\tilde{m}_R$, for 
$\tilde{m}^2_{R \mu \tau} = 0.8 \cdot \tilde{m}^2_R$  
and four choices of   
the other relevant mass ratios:  
1) $M_1 =\tilde{m}_L= \tilde{m}_R$ (solid line); 
2) $M_1= \tilde{m}_R/3$, $\tilde{m}_L= \tilde{m}_R$ (dotted line); 
3) $M_1= \tilde{m}_R$, $ \tilde{m}_L= \tilde{m}_R/3$ (dashed line); 
4) $M_1=\tilde{m}_R$, $\tilde{m}_L= 3\tilde{m}_R$ (thin solid line).}
\label{f3}
\vskip -0.1cm
\end{figure}
exhibit a common behaviour\footnote{
This behaviour would be visible for all the curves if we had not cut 
the axes.} with respect 
to the ratio $|\mu|/\tilde{m}_L$ or $|\mu|/\tilde{m}_R$: 
for each curve there is a deep minimum which separates 
the right-side region,  
where the pure $\tilde{B}^0$ diagram dominates as  
that mass ratio  increases (diagram (a) for $({\rm LFV})_L$
and diagram (e) for  $({\rm LFV})_R$ in Fig.~\ref{f1}),
from the left-side one in which the Higgsino-gaugino diagrams dominate. 
The deep wells  for either  $|\Delta_{L}|^2$ or $|\Delta_R|^2$ are due 
to the destructive interference of the above mentioned diagrams. 
Notice that the interference would be constructive if the sign of 
$M_1$ were opposite to that of $M_2$.   

In the case of $({\rm LFV})_L$ we can see  that values of  
$|50\Delta_L|^2$ larger than  $\sim 5\times 10^{-4}$ 
are achieved both in the left and right ranges in Fig.~\ref{f2}. 
The example
with $\tilde{m}_R= \tilde{m}_L/3$ (dashed line) 
provides larger values in the range $|\mu|/\tilde{m}_L \gsim 3$  since 
the pure $\tilde{B}^0$ diagram is further enhanced by the smaller 
$\tilde{m}_R$.
In the case of $({\rm LFV})_R$, values of  
$|50\Delta_R|^2$ larger than  $\sim 5\times 10^{-4}$ can  be 
obtained for large values of $|\mu|/\tilde{m}_R$ (see Fig.~\ref{f3}). 
An enhancement appears for $\tilde{m}_L= \tilde{m}_R/3$ (dashed line), 
in analogy to the $({\rm LFV})_L$ example mentioned above.
On the other hand, in the left-side region the values of 
$|50\Delta_R|^2$ are smaller with respect to the analogous 
ones of $|50\Delta_L|^2$.
Indeed, in this range $|\Delta_R|^2$  is dominated by the 
$\tilde{H}$-$\tilde{B}$ diagram (proportional to $g'^{2}$), 
while $|\Delta_L|^2$ 
is dominated by the $\tilde{H}$-$\tilde{W}$  diagrams 
(proportional to $g^{2}$).

We now  make contact with the physical observable, i.e. 
the $ BR(\Phi^0\to \mu^+\tau^-)$ in (\ref{rphi}), and discuss 
the phenomenological implications.
We recall that the Higgs boson masses and 
the angle $\alpha$ in the coefficients $C_\Phi$ 
are also affected, through radiative 
corrections, by a set of MSSM parameters not  involved in the 
determination of $\Delta_L, \Delta_R$,  
such as the mass parameters of the squark-gluino sector 
(see e.g. \cite{rc} and references therein).
The latter parameters  indirectly affect also 
the $BR(\Phi^0\to \tau^+\tau^-)$ through radiative 
corrections to $BR(\Phi^0\to b \bar{b})$ (see e.g. \cite{all1,GHP}).
We do not make a definite choice of those parameters and only
outline some  general features of 
$ BR(\Phi^0\to \mu^+\tau^-)$ at large $\tanb$ 
and the prospects  for these decay channels   
at the Large Hadron Collider (LHC) and other colliders\footnote{
For recent discussions and references on supersymmetric Higgs phenomenology 
see also \cite{higgs}. An unconventional scenario has been recently discussed 
in \cite{BCEN}.}.
It is convenient to schematically separate the three Higgs 
bosons into two groups. 
The CP-odd and one of the CP-even Higgs bosons have about 
the same mass, non-standard 
(enhanced) couplings with down-type fermions and suppressed couplings 
with up-type fermions and electroweak gauge bosons.
These bosons, which are mainly contained  in $H^0_1$, correspond 
to  $H^0, A^0$ ($h^0, A^0$) 
for $m_A \gsim m_\star$ ($m_A \lsim m_\star$), where  
$m_\star \sim 110 - 130~{\rm GeV}$.
The other CP-even Higgs  has a mass $\sim  m_\star$   
and Standard Model-like couplings with up-type fermions and 
electroweak gauge bosons.
It is mainly  contained in $H^0_2$ and   corresponds to 
$h^0$ ($H^0$) for  $m_A \gsim m_\star$ ($m_A \lsim m_\star$). 
Let us discuss the different Higgs bosons, assuming 
for definiteness $\tan\beta\sim 50$, 
$|50\Delta|^2 \sim 10^{-3}$ ($\Delta = \Delta_L$ or $\Delta_R$) and 
an integrated luminosity of $100 ~{\rm fb}^{-1}$ at LHC.
\begin{enumerate}
\item
If $\Phi^0$ denotes one of 
the `non-standard' Higgs bosons, we have $C_\Phi \simeq 1$ and 
$ BR(\Phi^0\to \tau^+\tau^-) \sim 10^{-1}$, so  
$ BR(\Phi^0\to \mu^+\tau^-) \sim 10^{-4}$. 
The main production mechanisms at LHC are  bottom-loop mediated gluon fusion 
and associated production with $b \bar{b}$, which yield cross sections 
 $\sigma\sim  (10^3, 10^2, 20)~{\rm pb}$ 
for $m_A \sim(100,200,300)~{\rm GeV}$, respectively. 
The corresponding numbers of $\Phi^0\to \mu^+\tau^-$ events are about  
$(10^4,10^3, 2\cdot 10^2)$.
These estimates do not change much if 
the bottom Yukawa coupling $Y_b$ is enhanced (suppressed)  
by  radiative corrections, 
since in this case the enhancement (suppression) of $\sigma$ would be  
roughly compensated by the suppression (enhancement) of   
$BR(\Phi^0\to \mu^+\tau^-)$. 
\item
If $\Phi^0$ denotes  the other (more `Standard Model-like') Higgs boson, 
the factor  $C_\Phi\cdot 
BR(\Phi^0\to \tau^+\tau^-)$  strongly depends on $m_A$, while  
the production cross section at LHC, which is 
dominated by top-loop mediated gluon fusion, is 
$\sigma \sim 30~{\rm pb}$.
For  $m_A\sim 100~{\rm GeV}$ we may have 
$C_\Phi\cdot BR(\Phi^0\to \tau^+\tau^-)\sim 10^{-1}$ and 
$BR(\Phi^0\to \mu^+\tau^-) \sim 10^{-4}$, which  
would imply $\sim 300$ $ \mu^+\tau^-$ events. 
The number of events is generically smaller 
for large   $m_A$ since  $C_\Phi$ scales as  $ 1/m_A^4$, consistently 
with the expected decoupling of LFV effects for such a Higgs boson.
However, an enhancement can occur under certain conditions. 
In particular, for a range of $m_A$ values  the (radiatively corrected) 
off-diagonal element of the Higgs boson mass matrix could be 
over-suppressed. 
In this case the $\Phi^0 b {b}^c,~ \Phi^0 \tau {\tau}^c$ 
couplings would also be suppressed and as a result   
the number of  $ \mu^+\tau^-$ events could be even  ${\cal O}(10^3)$.
\end{enumerate}

The above discussion suggests  that LHC may offer good chances 
to detect the decays $\Phi^0\to \mu\tau$, especially in the case
of non-standard Higgs bosons. 
This indication should be supported by a detailed study of the 
background (which is beyond the scope of this work),  for instance 
by generalizing  the analyses in \cite{gen}. 
At Tevatron the sensitivity is lower  than at LHC because 
both  the expected luminosity  
and  the Higgs production cross sections are  smaller. 
The number of events would be smaller by a factor $10^2 - 10^3$. 
A few events may be expected
also at $e^+ e^-$ or $\mu^+ \mu^-$ future  colliders,  
assuming integrated luminosities of 
$500~{\rm fb}^{-1}$ and $1~{\rm fb}^{-1}$, respectively.  
At a $\mu^+ \mu^-$  collider 
an enhancement  may occur for the non-standard Higgs bosons   
if radiative corrections strongly suppress $Y_b$, since in this
case both the resonant production cross section [$\sigma 
\sim (4\pi/m^2_A) BR( \Phi^0 \to \mu^+\mu^-)$] and the 
LFV branching ratios 
$BR(\Phi^0\to \mu^+\tau^-)$ would be enhanced.  
As a result, for 
light $m_A$, hundreds of  $\mu^+\tau^-$ events could occur. 

Finally we notice that, although we have focused on LFV decays 
of neutral Higgs bosons, 
also  charged Higgs bosons have LFV decays, i.e 
$   H^+ \to \tau^+ \nu_\mu$ and $H^+ \to \mu^+ \nu_\tau$ 
(and related charge conjugated channels). 
Also these decays are  controlled by the parameters 
$\Delta_L$ and $\Delta_R$, at lowest order in $SU(2)_W$       
breaking effects.
The FV couplings with the charged Higgs bosons 
emerge by  taking into account the $SU(2)_W$ completion 
of eqs.~(\ref{Ltree}) and (\ref{leff}).     
It is straightforward to find 
$BR(H^+\to \tau^+\nu_\mu)
=  \tan^2\beta~ |\Delta_L|^2 
 ~  { BR(H^+ \to \tau^+ \nu_\tau)} $ and 
$BR(H^+\to \mu^+\nu_\tau)
=  \tan^2\beta~ |\Delta_R|^2 ~{ BR(H^+\to \tau^+ \nu_\tau)} $. However, 
it is more natural to compare $H^+\to \mu^+\nu_\tau$ with 
$ H^+\to \mu^+\nu_\mu$ so that:
\be{phi+}
BR(H^+\to \mu^+\nu_\tau)
\simeq \left(\frac{m_\tau}{m_\mu}\right)^2 \tan^2\beta~ |\Delta_R|^2 
 ~ BR(H^+\to \mu^+\nu_\mu) .  
\ee
For $\tan^2\beta ~|\Delta_R|^2 \sim 10^{-3}$ this would lead 
to a 30\% enhancement in the 
channel $H^+\to \mu^+$ + {\it missing energy}.

\section{Final remarks and conclusions}

A few comments are in order about possible correlations 
between  the decays  $\Phi^0\to \mu\tau$ and other LFV processes.
We have seen that non-negligible rates for $\Phi^0\to \mu\tau$ can 
only be obtained for large $\tan\beta$ and large LFV. 
In this limit also the decay rate for $\tau \to \mu \gamma$,
which is dominated by diagrams analogous to those of Fig.~\ref{f1} 
with an extra photon attached \cite{lfv}, is enhanced and could exceed 
the experimental limit. 
However, we recall that the rate of $\tau \to \mu \gamma$ decreases 
as the superparticle masses increase, whereas 
the rate of $\Phi^0\to \mu\tau$ does not, since the 
latter is induced by dimension-four effective operators and only 
depends on mass ratios. Therefore to obtain an adequate suppression 
of $\tau\to \mu \ga$ the superparticle spectrum has to be raised  
towards the TeV region, although some slepton may be lighter.
For instance, in the case 1) of  $({\rm LFV})_L$ shown in Fig.~\ref{f2} 
($M_1= M_2 =\tilde{m}_R= \tilde{m}_L$),   
for $|\mu|/\tilde{m}_L \sim 1$ we obtain 
$|50 \Delta_L|^2 \sim 6\times 10^{-4}$.
In this particular example the present bound 
$BR(\tau \to \mu \gamma) < 6\times 10^{-7}$ \cite{belle1} 
constrains  $\tilde{m}_L \gsim 1.4~ {\rm TeV}$ 
for $\tan\beta=50$, which implies
${\rm min}(\tilde{m}_{L_2},\tilde{m}_{L_3}) \gsim 0.6~ {\rm TeV}$, 
${\rm max}(\tilde{m}_{L_2},\tilde{m}_{L_3}) \gsim 1.9~ {\rm TeV}$ and   
$M_1, M_2, \tilde{m}_R, |\mu| \gsim 1.4 ~{\rm TeV}$. 

The decays  $\Phi^0\to \mu\tau$ are also correlated to 
the decay $\tau \to 3\mu$. We recall that the latter
receives $\tanb$-enhanced contributions of two types: 
from dipole LFV operators via photon exchange \cite{lfv} 
and from the scalar LFV operators (\ref{LFV})  
via Higgs exchange \cite{BK,DER}.
The dipole contribution is directly related to the 
$\tau\to \mu \ga$ decay rate and is consequently bounded, i.e.  
$BR(\tau\to 3\mu)_{\ga^*} \sim 2.3\times 10^{-3} ~ BR( \tau\to \mu \ga) 
\lsim 1.4 \times 10^{-9}$. As for the Higgs-mediated contribution,
we obtain the  following estimate:  
\be{tau3}
BR(\tau\to 3\mu)_{\Phi^*} \sim 10^{-7} 
\left(
\frac{\tan\beta}{50}\right)^6  
\left(\frac{100 ~{\rm GeV}}{m_{A}}\right)^4 ~
\left( \frac{|50 \Delta_L|^2 +|50 \Delta_R|^2}{10^{-3}} \right)  ~.
\ee
Therefore, this contribution 
can exceed the dipole induced one\footnote{
Here our conclusion is in qualitative agreement with 
that drawn by \cite{BK}. 
On the other hand, the authors of \cite{DER} conclude that
Higgs-mediated contributions to  $\tau \to 3\mu$ are subleading
compared to the photonic penguin ones. 
We believe that this different conclusion is partly due to the fact 
that in \cite{DER} 
the superparticle masses are chosen to lie below the TeV scale, so
sizeable values for the LFV Higgs couplings are 
prevented by the $\tau\to \mu \ga$ constraint.} 
and be not far from 
the present bound, $BR(\tau \to 3 \mu) < 3.8 \times 10^{-7}$ \cite{belle2}.
Notice that the  parameter region in which this occurs  
is also the most favorable one for the observation of 
the $\Phi^0 \to \mu \tau$ decays, so an interesting correlation 
emerges.  
 
Throughout our work we have focused on the second and third generations, 
implicitly assuming that large slepton mixing only appears in
that sector. In a  scenario in which staus
are mainly mixed with selectrons rather than with smuons,
our discussion and numerical estimates concerning 
$\Phi^0 \to \mu \tau$ decays can be directly translated 
to $\Phi^0 \to e \tau$ decays, with obvious substitutions. 
The case of large smuon-selectron mixing is somewhat 
different. Although the strong constraints  
from $\mu \to e \gamma$ can be satisfied by taking sufficiently heavy 
superparticles, the decays $\Phi^0 \to \mu e $ 
are generically suppressed by the presence of $Y_\mu$.
The latter decays could be $Y_\tau$-enhanced if both 
$({\rm LFV})_L$ and $({\rm LFV})_R$ were present, 
and staus were mixed with both smuons and selectrons.

In summary, we have studied the LFV couplings of Higgs bosons  
in a general MSSM framework, allowing for generic LFV entries 
in the slepton mass matrices, but without invoking any specific 
mechanism to generate them.
We have computed the branching ratios 
of $\Phi^0 \to \mu \tau$ decays, which depend on ratios 
of MSSM mass parameters, and 
increase for increasing $\tanb$ and LFV.
Although cancellations can occur in some regions of parameter space,
${\cal O}(10^{-4})$ values 
 are achievable,  and they are compatible with the bounds 
on $\tau\to \mu \ga$ for a superparticle spectrum in the TeV range.
If the Higgs spectrum is relatively light ($m_A\lsim 300$ GeV), 
our results indicate that future colliders (in particular LHC) 
may be able to detect the decays $\Phi^0\to \mu\tau$, especially 
in the case of non-standard Higgs bosons. Moreover, the detection 
of these decays is closely correlated with that of $\tau \to 3 \mu$, 
which may be observed in the near future.

\vspace{0.9cm}

\noindent
{\bf Acknowledgments}\\

\noindent
This work  was   partially supported
by the European Union under the contracts 
HPRN-CT-2000-00148 (Across the Energy Frontier) and
HPRN-CT-2000-00149 (Collider Physics).

\end{document}